\documentclass[11pt]{article}
\title{Flexible Principal Component Analysis for Exponential Family Distributions}
\author{Tonglin Zhang \footnote{Department of Statistics, Purdue University, 250 North University Street,West Lafayette, IN 47907-2066, USA, Email: tlzhang@purdue.edu}, Baijian Yang \footnote{Department of Computer and Information Technology, Purdue University, 401 North Grant Street, West Lafayette, IN 47907, USA, Email: byang@purdue.edu},  Qianqian Song \footnote{Department of Cancer Biology, Wake Forest School of Medicine, One Medical Center Boulevard, Winston-Salem, NC 27157-1082, USA, Email: qsong@wakehealth.edu}, and Jing Su \footnote{Department of Biostatistics and Health Data Science, Indiana University School of Medicine, Indianapolis, IN 46202-3002, USA, Email: su1@iu.edu},}\usepackage{epsfig}
\usepackage{mathrsfs}
\usepackage{amsfonts}
\usepackage{amsmath}
\usepackage{wrapfig,lipsum,booktabs}
\usepackage{bm}
\usepackage{algorithm}
\usepackage{algpseudocode}
\usepackage{graphicx}
\usepackage{subcaption}
\def\qed{\hfill$\diamondsuit$}

\newtheorem{defn}{Definition}

\newtheorem{thm}{Theorem}
\newtheorem{cor}{Corollary}
\newtheorem{lem}{Lemma}
\begin{document}
\maketitle
\def\eqalign#1{\null\,\vcenter{\openup\jot\ialign
              {\strut\hfil$\displaystyle{##}$&$\displaystyle{{}##}$
               \hfil\crcr#1\crcr}}\,}


\begin{abstract}

Traditional principal component analysis (PCA) is well known in high-dimensional data analysis, but it requires to express data by a matrix with observations to be continuous. To overcome the limitations, a new method called flexible PCA (FPCA) for exponential family distributions is proposed. The goal is to ensure that it can be implemented to arbitrary shaped region for either count or continuous observations. The methodology of FPCA is developed under the framework of generalized linear models. It provides  statistical models for FPCA not limited to matrix expressions of the data. A maximum likelihood approach is proposed to derive the decomposition when the number of principal components (PCs) is known. This naturally induces a penalized likelihood approach to determine the number of PCs when it is unknown. By modifying it for missing data problems, the proposed method is compared with previous PCA methods for missing data. The simulation study shows that the performance of FPCA is always better than its competitors. The application uses the proposed method to reduce the dimensionality of arbitrary shaped sub-regions of images and the global spread patterns of COVID-19 under normal and Poisson distributions, respectively.
\end{abstract}



\section{Introduction}
\label{sec:introduction}

Principal component analysis (PCA) is a well known dimension reduction technique in high-dimensional data analysis if continuous observations can be expressed by a data matrix. Previous PCA methods rely on a matrix technique called singular value decomposition (SVD). SVD decomposes the data matrix into the product of a left matrix for the left singular vectors, a diagonal matrix for singular values, and a right matrix for the right singular vectors. By removing small singular values with corresponding singular vectors from the output, a low-rank approximation of the input matrix, called traditional PCA, is derived. We have identified at least two limitations in this method. The first is that traditional PCA cannot be used if observations cannot be expressed by a data matrix. The second is that traditional PCA cannot be used even if observations can expressed by a data matrix but all of the observations are count. The goal of the research is to develop a new method, called flexible PCA (FPCA), to overcome the two limitations.

\begin{figure}
\centerline{\rotatebox{0}{\psfig{figure=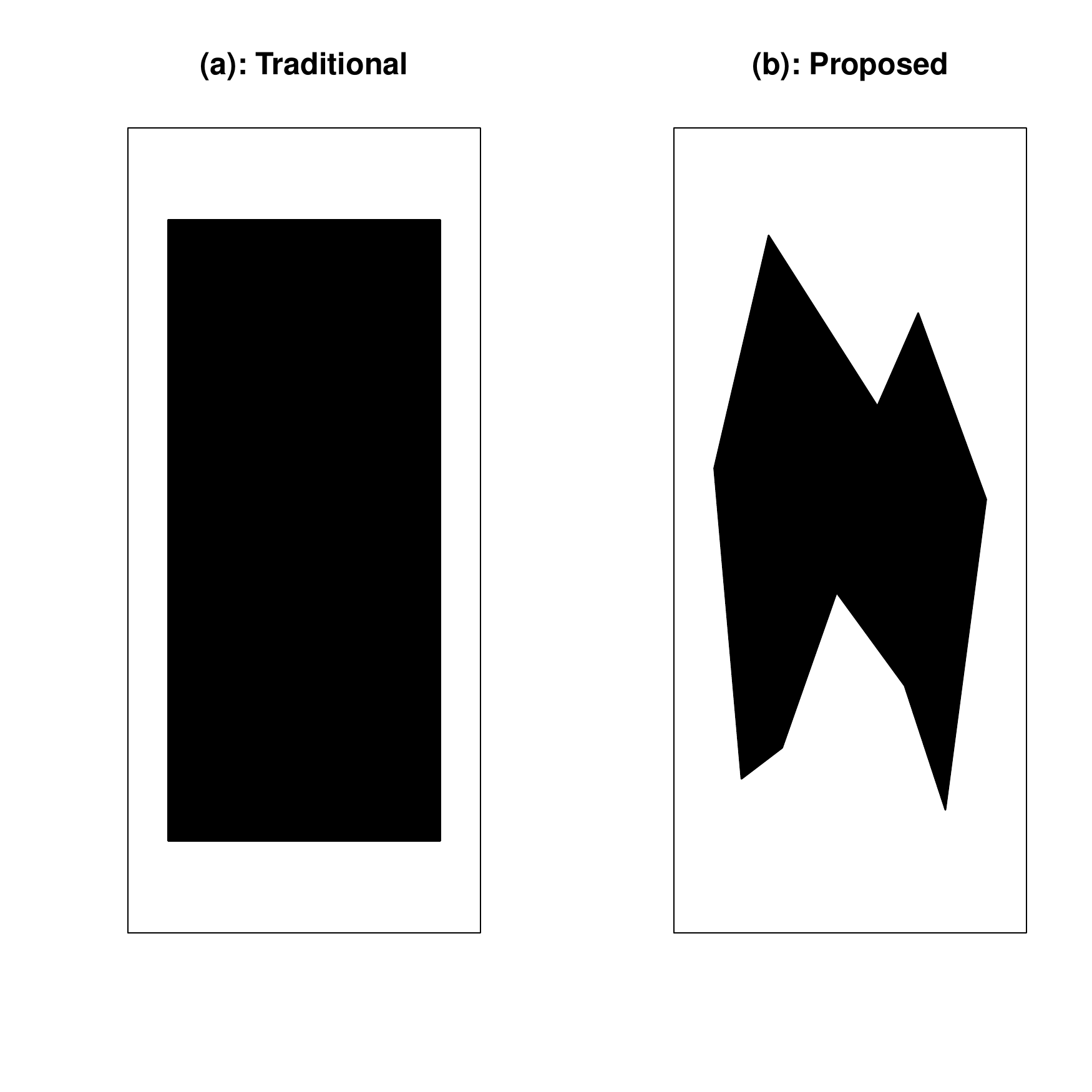,width=3.5in,}}}
\caption{\label{fig:pca region shape problem}Shapes of regions can only be considered by traditional PCA for matrices (the left), but our proposed FPCA can consider arbitrary shaped regions (the right).}
\end{figure}

We emphasize the contribution of our method by Figure~\ref{fig:pca region shape problem}. If an image data has been loaded by a computer and the observations have been expressed by a matrix (Figure~\ref{fig:pca region shape problem}(a)), then traditional PCA can be implemented to the data. This is a widely used approach in image analysis because any black-and-white image can be treated as a matrix and any color image can be treated as a combination of three matrices for RGB channels, respectively. However, if the interest is not the entire image but only a specific region contained by the image, then it is inappropriate to use traditional PCA because irrelevant information contained by pixels outside the region is involved. To overcome the difficulty, dimension reduction for arbitrary shaped regions is needed (Figure~\ref{fig:pca region shape problem}(b)). In addition, if all of the entries of the matrix are count, then it is inappropriate to use traditional PCA in the case displayed by Figure~\ref{fig:pca region shape problem}(a) either. Therefore, we need to address two challenges. The first is to ensure that FPCA can be implemented to arbitrary shaped regions. The second is to ensure that it can be implemented to count data.

We avoid SVD in our method. Instead, we use the maximum likelihood approach, where we need a statistical model. The model is obtained by extensions of previous PCA models for matrices. Three PCA models have been proposed in the literature. The first is the low-rank mean model~\cite[P. 146]{anderson2003}. The second is the latent factor model~\cite{lawrence2005,tipping1999}. The third is the spiked covariance model~\cite{johnstone2001}. We devise our model by an extension of the low-rank mean model because the latent factor and the spiked covariance models contain matrix transformations, not easy to interpret for arbitrary shaped regions. In addition, the low-rank mean model has been extended to count matrices by the generalized linear models (GLMs)~\cite{collins2001},  leading to a method called generalized PCA~\cite{landgraf2020}. Based on the maximum likelihood approach in the low-rank mean model for arbitrary shaped regions, we are able to address the two challenges that we have mentioned in the previous paragraph.

Our method is fundamentally different from previous PCA methods for missing values~\cite{dray2015,josse2016}, including generalized PCA~\cite{landgraf2020}. We purposely discard observations outside of the region of interest even though they are available. Therefore, it is not a missing data problem. The previous methods assume that data are given by matrix forms. Missing values appear if some entries of the matrix are not available. Imputations for missing values are needed because they use SVD for the completed data to derive the decomposition. This often involves the mechanisms for generating missing data in the evaluation of their theoretical properties. However, missing data mechanism is not an issue and SVD is purposely avoided in our method. 

In the literature, PCA~\cite{jolliffe2002} is a widely used dimension reduction technique for matrices data. Besides PCA, another often used technique is sufficient dimension reduction~\cite{cook2005,li2007}. The difference is that PCA does not need a response variable, but this is required in sufficient dimension reduction. Thus, PCA is an unsupervised learning method, but sufficient dimension reduction is not. Based on SVD for an input matrix, PCA can reduce the dimensionality of high-dimensional vector-valued variables of interest and simultaneously preserves their relationship~\cite{cook1994,li1989}. In the past a few decades, statistical approaches to dimension reduction have gained considerable attention due to rapid increases of data volume and dimension~\cite{cook2007,li2007,xia2002}. The implementation of dimension reduction techniques, including PCA, becomes popular in high-dimensional data analysis. Examples include dimension reduction in linear models~\cite{sousa2007}, generalized linear models (GLMs)~\cite{marx1990}, discriminant analysis~\cite{choi2004,zhao2000}, cluster analysis~\cite{gardner1991}, image analysis~\cite{celik2009,huangliu2020,zhang2010}, and big data~\cite{zhangyang2016,zhangyang2018}. All of these methods assume that observations can be expressed by  a matrix. PCA is an approximate method. It provides the best approximation based on the measure given by the Frobenius norm loss between the input matrix and the decomposition. This property naturally provides a regression model to interpret PCA. In FPCA, we use the maximum likelihood approach, including the least squares for the regression model, in the derivation. Therefore, our method is more flexible than previous PCA methods based on SVD.

Our method also includes the penalized likelihood approach for the determination of numbers of principal components (PCs). It recommends using the maximum likelihood approach to estimate the decomposition if the number of PCs is known. If the number of PCs is unknown, then we use the penalized likelihood approach. The development of the penalized likelihood approach is straightforward because the likelihood function is provided in each candidate models. In the literature, the determination of the number of PCs is considered as one of the most important problems in traditional PCA for matrices. This problem can be assessed by the penalized likelihood approach. By studying a few well known options of the tuning parameter in the penalized likelihood approach for variable selection, we conclude that the number of PCs can be determined by BIC but not AIC. This is consistent with a previous finding for the same problem  when data can be expressed by a matrix~\cite{bai2018}. Our research indicates that BIC can also be used to determine the number of PCs even if data cannot be expressed by a matrix.

The article is organized as follows. In Section~\ref{sec:method}, we introduce our method. In Section~\ref{sec:simulation}, by treating FPCA as a method for missing data, we compare our method with our competitors by simulations. The simulation results show that the performance of our method is always better than our competitors. In Section~\ref{sec:application}, we implement our method to three real world examples. Two are image analysis problems. One is an infectious disease problem. It is dimension reduction for the outbreak of COVID-19 in the world. In Section~\ref{sec:discussion}, we provide a discussion.

\section{Method}
\label{sec:method}

We propose statistical models for FPCA in Section~\ref{subsec:statistical model for FPCA}. We propose maximum likelihood estimation to derive the FPCA decomposition based on a selected number of PCs in Section~\ref{subsec:maximum likelihood estimation}. Because the determination of the number of PCs is important in PCA even for matrices, we propose a penalized likelihood approach to estimate the number of PCs in Section~\ref{subsec:penalized likelihood for number of PCs}. Although our method is not proposed for missing data problem, it can be modified for such a problem. This is introduced in Section~\ref{subsec:modification for pca with missing data}.

\subsection{Statistical Model for FPCA}
\label{subsec:statistical model for FPCA}

We propose our statistical model for FPCA by modifications of previous models for matrices. In the literature, three PCA models have been proposed for matrices. All of them assume that the number of PCs, denoted by $k$, is known. The first is the low-rank mean model~\cite{anderson2003}. It assumes that the data matrix is equal to the sum of the low-rank matrix and a white noise error matrix. The second is the latent factor model~\cite{lawrence2005,tipping1999}. It assumes that the data matrix is derived by a common transformation on a number of iid multivariate normal latent random factors. The third is the spiked covariance model~\cite{johnstone2001}. It assumes that singulars values contained by the PCA expression are higher than those not contained. The singular values are assumed to be constants in the low-rank mean and the latent factor models but not in the spiked covariance model. 

We propose our FPCA model by a modification of the low-rank mean model. The low-rank mean model is proposed for normally distributed $x_{ij}$ expressed by an $n\times p$ matrix with $i=1,\dots,n$ and $j=1,\dots,p$, such that
\begin{equation}
\label{eq:low-rank PCA model for matrix}
x_{ij}=\sum_{r=1}^k d_ru_{ir}v_{jr}+\epsilon_{ij}, \epsilon_{ij}\sim^{iid}{\cal N}(0,\sigma^2),
\end{equation}
where $d_r$, $u_{ir}$, and $u_{jr}$ for $r=1,\dots,k$ are parameters. To be consistent with the format of traditional PCA,  singular values $d_1,\dots,d_k$ are assumed to be positive and given by decreasing orders, ${\bm u}_1,\dots,{\bm u}_k\in\mathbb{R}^n$ are unit orthogonal vectors,  and ${\bm v}_1,\dots,{\bm v}_k\in\mathbb{R}^p$ are also  unit orthogonal vectors,  where ${\bm u}_r=(u_{1r},\dots, u_{nr})^\top$ and ${\bm v}_r=(v_{1r},\dots,v_{pr})^\top$ represent the $r$th left and right singular vectors, respectively.

Our interest is not the decomposition of the entire matrix but only the decomposition of those $x_{ij}\in {\cal S}$, where ${\cal S}$ is a subset of the matrix. We assume that the distribution of $x_{ij}$ follows an exponential family distribution with its probability density function (PDF) or probability mass function (PMF) as 
\begin{equation}
\label{eq:exponential family distribution}
f(x_{ij})=\exp\left[{x_{ij}\omega_{ij}-b(\omega_{ij})\over a(\phi_{ij})}+c(x_{ij},\phi_{ij})\right],
\end{equation}
where $\omega_{ij}$ is the canonical parameter and $\phi_{ij}$ is the dispersion parameter. Then, we have $\mu_{ij}={\rm E}(x_{ij})=b'(\omega_{ij})$ and ${\rm var}(x_{ij})=a(\phi_{ij})b''(\omega_{ij})$. The exponential family distribution given by~\eqref{eq:exponential family distribution} includes the normal, binomial, and Poisson distributions as its special cases. The dispersion parameter $\phi_{ij}$ is present in normal distributions but not in binomial or Poisson distributions. We assume that the dispersion parameter depends on $(i,j)$ because we want to use it to define  correlation and covariance FPCA in our method.

Based on the template of the low-rank mean model, we propose the general form of our FPCA model as
\begin{equation}
\label{eq:generalized PCA for exponential family distributions}
g[b'(\omega_{ij})]=g(\mu_{ij})=\gamma_{ij}+\sum_{r=1}^k \alpha_{ir}\beta_{ir},  (i,j)\in{\cal S},
\end{equation}
where $\gamma_{ij}$ will be further specified according to the specific FPCA models. In the simplest case, we assume that $\gamma_{ij}$ is absent, leading to a model with the second term on the right-hand side of~\eqref{eq:generalized PCA for exponential family distributions} only. In addition, we may choose $\gamma_{ij}=\gamma_j$ with either $\phi_{ij}=\phi_j$ or $\phi_{ij}=\phi$ in~\eqref{eq:exponential family distribution}, leading to the correlation FPCA or the covariance FPCA models. This is not involved in~\eqref{eq:low-rank PCA model for matrix} if $x_{ij}$ is not standardized. We discuss this issue in Section~\ref{subsec:maximum likelihood estimation}.

Although the second term on the right-hand side of~\eqref{eq:generalized PCA for exponential family distributions} is different from the first term on the right-hand side of~\eqref{eq:low-rank PCA model for matrix}, they are equivalent in the formulations of decomposition. This has been previously proven by many authors for matrices~\cite[e.g.]{collins2001,kiers1997}. The conclusion can be easily extended to any shaped ${\cal S}$. If ${\cal S}$ is the entire matrix, then $d_r$, $u_{ir}$ and $v_{jr}$ are unique in~\eqref{eq:low-rank PCA model for matrix}, implying that they are identifiable. This conclusion is violated if ${\cal S}$ is not the entire matrix~\cite{dray2015}. In~\eqref{eq:generalized PCA for exponential family distributions},  $\alpha_{ir}$ and $\beta_{jr}$ cannot be uniquely defined even if ${\cal S}$ is the entire matrix. Therefore, we need to evalaute the impact of the identifiability in both~\eqref{eq:low-rank PCA model for matrix} and~\eqref{eq:generalized PCA for exponential family distributions}. This can be addressed by model predictions.  We use this idea to develop  the maximum likelihood approach for~\eqref{eq:generalized PCA for exponential family distributions} in Section~\ref{subsec:maximum likelihood estimation}.

\subsection{Maximum Likelihood Estimation}
\label{subsec:maximum likelihood estimation}

We use maximum likelihood estimation to compute the decomposition given by~\eqref{eq:generalized PCA for exponential family distributions} with a selected $k$. Let ${\bm\alpha}$ be the parameter vector composed by $\alpha_{ir}$, ${\bm\beta}$ be that composed by $\beta_{jr}$, and ${\bm\gamma}$ and ${\bm\phi}$ be those composed by $\phi_{ij}$ and $\gamma_{ij}$ for all $(i,j)\in{\cal S}$ and $r=1,\dots,k$, respectively. The loglikelihood function of the model jointly defined by~\eqref{eq:exponential family distribution} and~\eqref{eq:generalized PCA for exponential family distributions} is
\begin{equation}
\label{eq:loglikelihood function of gpca}
\ell_k({\bm\theta},{\bm\phi})=\sum_{(i,j)\in {\cal S}} \left\{{x_{ij}h(\gamma_{ij}+\sum_{r=1}^k \alpha_{ir}\beta_{jr})-b[h(\gamma_{ij}+\sum_{r=1}^k \alpha_{ir}\beta_{jr})]\over a(\phi_{ij})} +c(x_{ij},\phi_{ij}) \right\},
\end{equation}
where ${\bm\theta}=({\bm\alpha}^\top,{\bm\beta}^\top,{\bm\gamma}^\top)^\top$ and $h(\cdot)$ is the inverse function of~\eqref{eq:generalized PCA for exponential family distributions} defined by $\omega_{ij}=h(\gamma_{ij}+\sum_{r=1}^k\alpha_{ir}\beta_{jr})$. The MLE of ${\bm\theta}$ and ${\bm\phi}$ given $k$, denoted by $\hat{\bm\theta}_k$ and $\hat {\bm\phi}_k$, respectively, is 
\begin{equation}
\label{eq:MLE of theta given k}
(\hat{\bm\theta}_k^\top,\hat{\bm\phi}_k^\top)^\top=(\hat{\bm\alpha}_k^\top,\hat{\bm\beta}_k^\top,\hat{\bm\gamma}_k^\top,\hat{\bm\phi}_k^\top)^\top=\mathop{\arg\!\min}_{\bm\theta}\ell_k({\bm\theta},{\bm\phi}).
\end{equation}

The loglikelihood function given by~\eqref{eq:loglikelihood function of gpca} is too general. In practice, we recommend using its three special cases. The first assumes that $\gamma_{ij}$ is absent in~\eqref{eq:generalized PCA for exponential family distributions} and $\phi_{ij}=\phi$ are all the same in~\eqref{eq:exponential family distribution}, leading to the first modification of~\eqref{eq:loglikelihood function of gpca} as
\begin{equation}
\label{eq:modified loglikelihood function of gpca original}
\ell_{1,k}({\bm\theta},\phi)=\sum_{(i,j)\in {\cal S}} \left\{{x_{ij}h(\sum_{r=1}^k \alpha_{ir}\beta_{jr})-b[h(\sum_{r=1}^k \alpha_{ir}\beta_{jr})]\over a(\phi)} +c(x_{ij},\phi) \right\}.
\end{equation}
In this case, we denote the solution of~\eqref{eq:MLE of theta given k} as $\hat{\bm\theta}_{1,k}=(\hat{\bm\alpha}_{1,k}^\top,\hat{\bm\beta}_{1,k}^\top,\hat{\bm\gamma}_{1,k}^\top)^\top$ for $\hat{\bm\theta}_k$ and $\hat\phi_{1,k}$ for $\hat{\bm\phi}_k$. The second assumes that $\gamma_{ij}=\gamma_j$ only depends on $j$ and $\phi_{ij}=\phi$ are all the same, leading to the second modification of~\eqref{eq:loglikelihood function of gpca} as
\begin{equation}
\label{eq:modified loglikelihood function of gpca covariance}
\ell_{2,k}({\bm\theta},\phi)=\sum_{(i,j)\in {\cal S}} \left\{{x_{ij}h(\gamma_j+\sum_{r=1}^k \alpha_{ir}\beta_{jr})-b[h(\gamma_j+\sum_{r=1}^k \alpha_{ir}\beta_{jr})]\over a(\phi)} +c(x_{ij},\phi) \right\}.
\end{equation}
In this case, we denote the solution of~\eqref{eq:MLE of theta given k} as $\hat{\bm\theta}_{2,k}=(\hat{\bm\alpha}_{2,k}^\top,\hat{\bm\beta}_{2,k}^\top,\hat{\bm\gamma}_{2,k}^\top)^\top$ for $\hat{\bm\theta}$ and $\hat\phi_{2,k}$ for $\hat{\bm\phi}_k$. The third assumes that $\gamma_{ij}=\gamma_j$ and $\phi_{ij}=\phi_j$ depend on $j$ only, leading to the third modification of~\eqref{eq:loglikelihood function of gpca} as
\begin{equation}
\label{eq:modified loglikelihood function of gpca correlation}
\ell_{3,k}({\bm\theta},{\bm\phi})=\sum_{(i,j)\in {\cal S}} \left\{{x_{ij}h(\gamma_j+\sum_{r=1}^k \alpha_{ir}\beta_{jr})-b[h(\gamma_j+\sum_{r=1}^k \alpha_{ir}\beta_{jr})]\over a(\phi_j)} +c(x_{ij},\phi_j) \right\},
\end{equation} 
where ${\bm\phi}=(\phi_1,\dots,\phi_p)^\top$.  In this case, we denote the solution of~\eqref{eq:MLE of theta given k} as $\hat{\bm\theta}_{3,k}=(\hat{\bm\alpha}_{3,k}^\top,\hat{\bm\beta}_{3,k}^\top,\hat{\bm\gamma}_{3,k}^\top)^\top$ for $\hat{\bm\theta}_k$ and $\hat{\bm\phi}_{3,k}$ for $\hat{\bm\phi}_k$. Then, we have the following theorems.

\begin{lem}
\label{lem:covariance and correlation PCA normal}
If $f(x_{ij})$ given by~\eqref{eq:exponential family distribution} is normal PDF and ${\cal S}$ is the entire matrix, then the method for $\hat{\bm\theta}_{2,k}$ and $\hat\phi_{2,k}$ is equivalent to covariance PCA, and the method for $\hat{\bm\theta}_{3,k}$ and $\hat{\bm\phi}_{3,k}$ is equivalent to correlation PCA.
\end{lem}

\noindent
{\bf Proof.} The proof is straightforward. We simply compare~\eqref{eq:modified loglikelihood function of gpca covariance} and~\eqref{eq:modified loglikelihood function of gpca correlation} with the objective functions in covariance and correlation PCA, respectively. We find that the objective functions equivalent, respectively. We draw the conclusion. \qed

\begin{defn}
\label{defn:generalized covariance and correlation PCA models}
The PCA model jointly defined by~\eqref{eq:low-rank PCA model for matrix} and~\eqref{eq:generalized PCA for exponential family distributions} with $\gamma_{ij}=0$ and $\phi_{ij}=\phi$ is called the simple FPCA model. The PCA model jointly defined by~\eqref{eq:low-rank PCA model for matrix} and~\eqref{eq:generalized PCA for exponential family distributions} with $\gamma_{ij}=\gamma_j$ and $\phi_{ij}=\phi$ is called the covariance FPCA model. The PCA model jointly defined by~\eqref{eq:low-rank PCA model for matrix} and~\eqref{eq:generalized PCA for exponential family distributions} with $\gamma_{ij}=\gamma_j$ and $\phi_{ij}=\phi_j$ is called the correlation FPCA model. 
\end{defn}

\begin{cor}
\label{cor:solutions to generalized covariance and correlation PCA models}
For any ${\cal S}$ and $k$,  the solution of the simple FPCA model is $\hat{\bm\theta}_{1,k}$ and $\hat\phi_{1,k}$, the solution of the covariance FPCA model is $\hat{\bm\theta}_{2,k}$ and $\hat\phi_{2,k}$, and the solution of the correlation FPCA model is $\hat{\bm\theta}_{3,k}$ and $\hat{\bm\phi}_{3,k}$, derived by maximizing~\eqref{eq:modified loglikelihood function of gpca original}, \eqref{eq:modified loglikelihood function of gpca covariance} and~\eqref{eq:modified loglikelihood function of gpca correlation}, respectively.
\end{cor}

\noindent
{\bf Proof.} The conclusion is directly implied by Definition~\ref{defn:generalized covariance and correlation PCA models}. \qed

\bigskip
We next provide a numerical algorithm to compute $(\hat{\bm\theta}_{1,k}^\top,\hat\phi_{1,k})^\top$, $(\hat{\bm\theta}_{2,k}^\top,\hat\phi_{2,k}^\top)$, and $(\hat{\bm\theta}_{3,k}^\top,\hat{\bm\phi}_{3,k}^\top)^\top$ in the simple, covariance, and correlation FPCA models, respectively. We cannot directly use the numerical algorithm for GLMs because the second term on the right-hand side of~\eqref{eq:generalized PCA for exponential family distributions} is not a linear function of unknown parameters.  We examine our model and find that it is connected with a previous method called generalized bilinear regression~\cite{gabriel1998}.  Then, we decide to use the generalized bilinear regression approach to fit our models. The main algorithm in generalized bilinear regression is the alternating GLM (AGLM). AGLM is an iterative algorithm. It provides the MLEs of generalized bilinear models for exponential family distributions. It becomes the alternating least squares (ALS) if the distribution is normal. As a special case, ALS is one of the most important algorithms in matrix and tensor decomposition problems~\cite{kolda2009,liu2020}. Examples include the CP-decomposition~\cite{carroll1970} and the Tucker decomposition~\cite{tucker1966} for tensors.

Our AGLM is composed by numerical algorithms for two working GLMs. The first working GLM treats ${\bm\beta}$ as constants, leading to a GLM for the rest parameters contained in ${\bm\theta}$ (i.e., but not ${\bm\beta}$) and ${\bm\phi}$. The second working GLM treats ${\bm\alpha}$ as constants, lead to a GLM for the rest parameters contained in ${\bm\theta}$ (i.e., but not ${\bm\alpha}$) and ${\bm\phi}$. Both can be fitted by the usual algorothms for GLMs. To start our AGLM, we need to initialize ${\bm\beta}$. To be consistent with the traditional ALS for tensor decomposition, we generate $\beta_{ij}^{(0)}$ independently from ${\cal N}(0,1)$. It provides the initial guess of $\hat{\bm\beta}_k$, denoted by ${\bm\beta}^{(0)}$. We do not need an initial guess of $\hat{\bm\alpha}_k$. Then, we have Algorithm~\ref{alg:AGLM for PCA fixed k}.

\begin{algorithm}
\caption{\label{alg:AGLM for PCA fixed k}AGLM for $\hat{\bm\theta}_k$ and $\hat{\bm\phi}_k$ based on a given $k$}
\begin{algorithmic}[1]
\Statex{{\bf Input}: $x_{ij}$ for all $(i,j)\in{\cal S}$}
\Statex{{\bf Output}: $\hat{\bm\theta}_{1,k}$ and $\hat\phi_{1,k}$, $\hat{\bm\theta}_{2,k}$ and $\hat\phi_{2,k}$, or $\hat{\bm\theta}_{3,k}$ and $\hat{\bm\phi}_{3,k}$,depending on whether it is the simple, covariance, or correlation FPCA model}
\Statex{\it Initialization}
\State{Independently generate $\beta_{jr}^{(0)}$ for $j=1,\dots,p$ and $r=1,\dots,k$ by ${\cal N}(0,1)$ and denote it as ${\bm\beta}^{(0)}$}
\Statex{\it Begin Iteration}
\State{${\bm\beta}\leftarrow{\bm\beta}^{(t-1)}$}
\State{Treat ${\bm\beta}$ as constants. Use the usual GLM algorithm to compute the MLE of all the rest parameters contained in ${\bm\theta}$ and ${\bm\phi}$. Obtain ${\bm\alpha}^{(t)}$, the $t$th iterated vector of $\hat{\bm\alpha}_k$.}
\State{${\bm\alpha}\leftarrow{\bm\alpha}^{(t)}$}
\State{Treat ${\bm\alpha}$ as constants. Use the same idea to derive ${\bm\beta}^{(t)}$, the $t$th iterated vector of $\hat{\bm\beta}_k$.}
\State{Repeat Steps 2-5 until convergence}
\Statex{\it End Iteration}
\State {Output}
\end{algorithmic}
\end{algorithm}

Because only the usual GLM algorithm is needed in the iterations, Algorithm~\ref{alg:AGLM for PCA fixed k} is efficient. It can be implemented even if $|{\cal S}|$ is extremely large. Although the final solution given by Algorithm~\ref{alg:AGLM for PCA fixed k} may not be the global maxmizer, we guarantee that it is one of the local maximizers. We state it as a theorem.

\begin{thm}
\label{thm:local optimizer PCA by ALS}
The final solution given by Algorithm~\ref{alg:AGLM for PCA fixed k} is a local maximizer. If there is only one local maximizer, then it is the global maximizer.
\end{thm}

\noindent
{\bf Proof.} Because the derivation of ${\bm\alpha}^{(t)}$ and ${\bm\beta}^{(t)}$ always makes the loglikelihood function higher, the final solution given by  Algorithm~\ref{alg:AGLM for PCA fixed k} cannot be a local minimizer or saddle point. It must be a local maximizer.  \qed

\bigskip
Although the final solution given by Algorithm~\ref{alg:AGLM for PCA fixed k} may not be the global maximizer, we can still use it to derive the global solution in the case when multiple local maximizers are present. The idea is to use multiple choices of ${\bm\beta}^{(0)}$. This is applicable because ${\bm\beta}^{(0)}$ is chosen by random. For all of those choices of ${\bm\beta}^{(0)}$, we calculate the corresponding values of the loglikelihood function. We report the one with the largest value as the final solution. We evaluated this approach by Monte Carlo simulations. We found that we were always able to obtain the global solution if five randomly generated ${\bm\beta}^{(0)}$ were used. Therefore, we are able to assume that the true $\hat{\bm\theta}_k$ and $\hat{\bm\phi}_k$ can always be obtained by Algorithm~\ref{alg:AGLM for PCA fixed k}.

\subsection{Penalized Likelihood for Number of PCs}
\label{subsec:penalized likelihood for number of PCs}

We devise a penalized likelihood approach to estimate $k$ if it is unknown. This is called the determination of the number of CPs in traditional PCA for matrices. It is important in the implementation of the method to real world data. We show that the problem can be addressed by the penalized likelihood approach. Penalized likelihood is a well accepted approach in variable selection. It has also been used in the determination of number of PCs in traditional PCA for matrices~\cite{bai2018}. Here, we adopt the well known GIC approach~\cite{zhangli2010} to construct our objective function with the best $k$ obtained by optimizing it.

We devise our penalized likelihood approach according to two scenarios. In the first, we assume that $\phi_{ij}$ is not present in~\eqref{eq:low-rank PCA model for matrix}. We can straightforwardly define the objective function in our GIC without any difficulty. This is applied to the case when $x_{ij}$ are binomial or Poisson random variables. In the second, we assume that $\phi_{ij}$ is present in~\eqref{eq:low-rank PCA model for matrix}. We need to provide an appropriate estimate of $\phi_{ij}$ suitable for all of the candidates of $k$. The common approach is to use a sufficiently large $k$ to derive the estimates and then fix them for all of the candidates of $k$~\cite{fanguo2012}. This is applied to the case when $x_{ij}$ are normal random variables. If $x_{ij}$ are binomial or Poisson random variables but overdispersion is present, then we also use the second scenario because we need $\phi_{ij}$ to interpret the overdispersion effect. 

To be consistent with the format of the objective functions in the traditional penalized likelihood approach for variable selection, we incorporate a tuning parameter, denoted by $\kappa$, in our method. For any candidate $k$, the maximizer of $\ell_k({\bm\theta},{\bm\phi})$ is achieved by $\ell_k(\hat{\bm\theta}_k,\hat{\bm\phi})$, where $\hat{\bm\phi}$ is the common estimator of ${\bm\phi}$ used for all candidates of $k$. Following~\cite{zhangli2010}, we define our GIC as
\begin{equation}
\label{eq:GIC objective function}
{\rm GIC}_{\kappa}(k)=-2\ell_k(\hat{\bm\theta}_k,\hat{\bm\phi})+\kappa df_k,
\end{equation}
where $\kappa$ is the tuning parameter that controls the properties of GIC and $df_k=k(n+p-k)$  is the degrees of freedom of the model. The best $k$ is solved by
\begin{equation}
\label{eq:the best k}
\hat k_{\kappa}=\mathop{\arg\!\min}_k\{{\rm GIC}_{\kappa}(k)\}.
\end{equation}
The GIC given by~\eqref{eq:GIC objective function} includes AIC if $\kappa=2$ or BIC if $\kappa=\log|{\cal S}|$. If they are adopted, then the solutions given by~\eqref{eq:the best k} are denoted by $\hat k_{AIC}$ and $\hat k_{BIC}$, respectively. Because AIC is inconsistent in estimating $k$ in traditional PCA for matrices~\cite{bai2018}, we discard $\hat k_{AIC}$ and focus on $\hat k_{BIC}$ in our method. We show that $\hat k_{BIC}$ is a consistent estimator of $k$ under $|{\cal S}|\rightarrow\infty$. 

\begin{thm}
\label{thm:consistent estimator of k}
Let $k_0$ be the true value of $k$. Assume that (i) $\ell_k({\bm\theta},{\bm\phi})$ given by~\eqref{eq:loglikelihood function of gpca} is specified to $\ell_{1,k}({\bm\theta},\phi)$, $\ell_{2,k}({\bm\theta},\phi)$, and $\ell_{1,k}({\bm\theta},{\bm\phi})$, given by~\eqref{eq:modified loglikelihood function of gpca original},~\eqref{eq:modified loglikelihood function of gpca covariance}, or~\eqref{eq:modified loglikelihood function of gpca correlation}, (ii) $\hat{\bm\phi}$ is a consistent estimator of ${\bm\phi}$, (iii) $n,p\rightarrow\infty$, $\kappa/\sqrt{2\log|{\cal S|}}\rightarrow\infty$, 
 $\kappa/\min\{n,p\}\rightarrow 0$, and $k_0/\min\{n,p\}\rightarrow 0$ as $|{\cal S}|\rightarrow\infty$, and (iv) $\mathop{\lim\!\inf}_{|{\cal S}|\rightarrow\infty}|{\cal S}|^{-1}\min\{d_1,\dots,d_{k_0}\}>0$ and  $\mathop{\lim\!\inf}_{|{\cal S}|\rightarrow\infty}|{\cal S}|/(np)>0$. Then $\hat k_{\kappa}-k_0\stackrel{P}\rightarrow0$ as $|{\cal S}|\rightarrow\infty$.
\end{thm}

\noindent
{\bf Proof.} Let ${\bm\phi}_0$ and ${\bm\theta}_0$  be the true ${\bm\phi}$ and ${\bm\theta}$, respectively.  Then, $\hat{\bm\phi}-{\bm\phi}_0\stackrel{P}\rightarrow 0$ by (ii). Following the standard proof for consistency of GIC estimators~\cite{zhangli2010}, we replace $\hat{\bm\phi}$ by ${\bm\phi}_0$ in~\eqref{eq:GIC objective function}. We want to show that the resulting estimator is consistent. If $\hat k_{\kappa}<k_0$, then there exists at least one of $r\in\{1,\dots,k_0\}$ such that the estimators of ${\bm\alpha}_r=(\alpha_{1r},\dots,\alpha_{nr})^\top$ and ${\bm\beta}_r=(\beta_{1r},\dots,\beta_{pr})^\top$ are not included in the final result of $\hat{\bm\theta}_k$. If $|{\cal S}|$ is sufficiently large,  then $2[\ell_k(\hat{\bm\theta}_k,\sigma_0^2)-\ell_k({\bm\theta}_0,\sigma_0^2)]$ approximately follows a non-central $\chi^2$-distribution with the magnitude of the non-centrality at least $[\min(n,p)\min\{d_1,\dots,d_{k_0}\}]^2$. The non-centrality disappears when $\hat k_{\kappa}$ reaches $k_0$. For any $k<k_0-1$, if we increase $k$ by $1$, then the change of the first term on the right-hand side of~\eqref{eq:GIC objective function} dominates that of the second, because $\kappa/\min\{n,p\}\rightarrow0$. Then, we have $P(\hat k_{\kappa}<k_0)\rightarrow 0$. If $\hat k_{\kappa}>k_0$, then the second term on the right-hand side of~\eqref{eq:generalized PCA for exponential family distributions} contains the true values of ${\bm\alpha}$ and ${\bm\beta}$. By the standard theory of the likelihood ratio test~\cite[Chapter 22]{ferguson1996}, we conclude that $2[\ell_k(\hat{\bm\theta}_k,{\bm\phi}_0)-\ell_k({\bm\theta}_0,{\bm\phi}_0)]$ approximately follows $\chi_{|{\cal S}|-df_k}^2$ when $|{\cal S}|$ is sufficiently large. For any $k>k_0+1$, if we decrease $k$ by $1$, then the change of the second term on the right-hand side of~\eqref{eq:GIC objective function} dominates that of the first, because $\kappa\rightarrow\infty$. As multiple candidates of $k$ are investigated, we need to adjust our condition by multiple testing problems. This can be addressed by the Higher Criticism approach~\cite{donoho2004}, which needs a stronger condition as $\kappa/\sqrt{2\log|{\cal S}|}\rightarrow\infty$ for $\kappa$. It is contained by (ii). Thus, we have $P(\hat k_{\kappa}>k_0)\rightarrow 0$ as $|{\cal S}|\rightarrow\infty$.  We replace ${\bm\phi}_0$ by $\hat{\bm\phi}$ in the expression of $\hat k_{\kappa}$ and obtain the final conclusion. \qed

\begin{cor}
\label{cor:consistent of BIC}
If all of the assumptions of Theorem~\eqref{thm:consistent estimator of k} hold, then $\hat k_{BIC}-k_0\stackrel{P}\rightarrow 0.$ 
\end{cor}

\noindent
{\bf Proof.} In BIC, we choose $\kappa=\log|{\cal S}|$. It satisfies assumption (iii) of Theorem~\ref{thm:consistent estimator of k}.  \qed

\bigskip
We cannot conclude consistency of $\hat k_{AIC}$ because it violates assumption (iii) of Theorem~\ref{thm:consistent estimator of k}. Therefore, we do not recommend using $\hat k_{AIC}$ in the determination of the number of PCs. We recommend using $\hat k_{BIC}$ by Corollary~\ref{cor:consistent of BIC}. BIC is not available in previous PCA methods for missing values because they also use the penalized likelihood to compute their PCA decomposition when $k$ is given. To compare our method with the previous methods, we need to devise a cross-validation approach to select the best $k$. This means that we need to modify our method for missing data problems.

\subsection{Modification for PCA with Missing Data}
\label{subsec:modification for pca with missing data}

Our method is different from previous PCA methods for missing values. To compare, we need to modify  our method for missing values. The modification is straightforward. We assume that ${\cal S}$ is not fixed but random. It is composed by randomly selected entries of the data matrix, such that  $x_{ij}$ for $(i,j)\in{\cal S}$ are observed and $x_{ij}$ for $(i,j)\not\in{\cal S}$ are missing. This is consistent with the settings of previous PCA methods for missing values as they are developed under a few well accepted missing data mechanisms~\cite{rubin1976}, such as missing completely at random (MCAR) or missing at random (MAR).  The missing data mechanisms are violated in the problems studied by Figure~\ref{fig:pca region shape problem}. Therefore, our method is fundamentally different from previous PCA methods for missing data. We can only compare the modification of our method with the previous methods. 

The maximum likelihood approach introduced in Section~\ref{subsec:maximum likelihood estimation} and the penalized likelihood approach introduced in Section~\ref{subsec:penalized likelihood for number of PCs} can be used even if ${\cal S}$ is random. Therefore, we can use BIC to estimate $k$ with the FPCA decomposition given by $\hat{\bm\alpha}_k$ and $\hat{\bm\beta}_k$. The implementation of our method does not need SVD in the computation of $\hat{\bm\theta}_k$, which can refer to $\hat{\bm\theta}_{1,k}$, $\hat{\bm\theta}_{2,k}$, or $\hat{\bm\theta}_{3,k}$. This is different from previous methods for PCA with missing values, as they either contain SVD for the completed data set derived by imputing the missing values~\cite{josse2016} or a penalty term to control orthogonality of the estimators of the factor matrices~\cite{landgraf2020}.  The previous methods propose their objective functions based on a weighted least squares criterion,  where the weight is one if the value of the entry is observed or zero otherwise. The derivation of the decomposition relies on an algorithm called iterative PCA or EM-PCA, where the EM algorithm is used to impute missing values and SVD is used to construct PCA. These methods cannot be combined with the penalized likelihood approach for the determination of $k$. Instead, they use the cross-validation approach to estimate $k$. To compare, we also propose a cross validation approach to estimate $k$ in our method. 

Because the computation of the leave-one-out cross-validation is time-consuming, we devise the training-testing cross-validation. We randomly partition ${\cal S}$ into a training data set ${\cal S}_{train}$ and a testing data set ${\cal S}_{test}$ with $P\{(i,j)\in{\cal S}_{train}\}=1-q$ and $P\{(i,j)\in{\cal S}_{test}\}=q$ for some $q\in(0,1)$. For each candidate $k$, we fit our PCA model defined by~\eqref{eq:exponential family distribution} and~\eqref{eq:generalized PCA for exponential family distributions} using observations in ${\cal S}_{train}$. We predict the values of $x_{ij}$, denoted by $\hat x_{ij}$, for all $(i,j)\in{\cal S}_{test}$. We calculate the difference between $x_{ij}$ and $\hat x_{ij}$ by the deviance $G_{{\cal S}_{test}}^2$ for the prediction. We choose $G_{{\cal S}_{test}}^2=\sum_{(i,j)\in{\cal S}_{test}}(x_{ij}-\hat x_{ij})^2$ if $f(x_{ij})$ given by~\eqref{eq:exponential family distribution} is a normal PDF or $G_{{\cal S}_{test}}^2$ as the usual deviance goodness-of-fit statistic if $f(x_{ij})$ is the binomial or Poisson PDF. We report $\hat k$, the estimate of $k$, by the lowest $G_{{\cal S}_{test}}^2$ value. An advantage is that the cross-validation approach does not contain a tuning parameter to be determined. An disadvantage is that the computation is usually time-consuming because the procedure for the partition of ${\cal S}$ into a training data set and a testing data set should be carried out multiple times to stabilize the results. This is a concern when $|{\cal S}|$ is large. In this case, we recommend using the penalized likelihood approach.

\section{Simulation}
\label{sec:simulation}

We compared our method with two previous PCA methods for missing data by simulations. The first was PCA for missing values~\cite{josse2016} carried out by \textsf{missMDA} package of \textsf{R}. The second was generalized PCA for exponential family distributions~\cite{landgraf2020} carried out by \textsf{generalizedPCA} package of \textsf{R}. If ${\cal S}$ is not a matrix, then \textsf{missMDA} and \textsf{generalizedPCA} treat it as a missing data problem but we treat it as an arbitrary shaped region problem. Therefore, we adopted the modification of our method for missing data developed Section~\ref{subsec:modification for pca with missing data} in the comparison. To be consistent with assumptions of \textsf{missMDA} and \textsf{generalizedPCA}, we  generated data from our covariance FPCA model as
\begin{equation}
\label{eq:data generation in simulation}
x_{ij}=\mu_j+\sum_{r=1}^k \alpha_{ir}\beta_{jr}+\epsilon_{ij}, \epsilon_{ij}\sim^{iid} {\cal N}(0,0.1^2),
\end{equation}
for all $i=1,\dots,n$ and $j=1,\dots,p$. We chose $n=p=30,60$ and $k=2,3,4$. We independently generated $\mu_j$ from ${\cal N}(0.5,4)$. For each $k$, we generated  $\alpha_{ir}$ and $\beta_{jr}$ independently from ${\cal N}(0,1)$. After all $x_{ij}$ were derived, we generated iid $Bernoulli(\tau)$ for each $x_{ij}$. We changed $x_{ij}$ to \textsf{NA} if the Bernoulli random value was $1$. Thus, the missing probability was $\tau$.

\begin{table}
\caption{\label{tab:determination of k our AIC and BIC}Percentage of correct estimates of $k$ by $\hat k_{AIC}$ and $\hat k_{BIC}$ in our FPCA based on simulations with $1000$ replications when data are generated from~\eqref{eq:data generation in simulation}.}
\begin{center}
\begin{tabular}{cccccccc}\hline
     &   & \multicolumn{2}{c}{$k=2$} & \multicolumn{2}{c}{$k=3$} &  \multicolumn{2}{c}{$k=4$}\\\cline{3-8}
$\tau$& $n$ & AIC & BIC & AIC & BIC & AIC & BIC \\\hline
$0.1$&$30$&$3.2$&$100.0$&$15.4$&$100.0$&$31.3$&$100.0$\\
&$60$&$14.6$&$100.0$&$12.7$&$100.0$&$26.3$&$100.0$\\
$0.2$&$30$&$0.0$&$100.0$&$5.4$&$100.0$&$27.1$&$100.0$\\
&$60$&$6.7$&$100.0$&$27.7$&$100.0$&$20.5$&$100.0$\\
$0.5$&$30$&$0.0$&$100.0$&$0.4$&$99.7$&$18.7$&$98.3$\\
&$60$&$0.0$&$100.0$&$1.9$&$100.0$&$19.6$&$100.0$\\\hline
\end{tabular}
\end{center}
\end{table}

We assumed that $k$ was unknown. We estimated $k$ by our $\hat k_{AIC}$ and $\hat k_{BIC}$ in our method. We found that $k$ was able to be estimated by $\hat k_{BIC}$ but not $\hat k_{AIC}$ (Table~\ref{tab:determination of k our AIC and BIC}). To compare, we also estimated $k$ by the training-testing cross-validation. To implement the training-testing cross-validation, we randomly partitioned observations into a training data set ${\cal S}_{train}$ and a test data set ${\cal S}_{test}$ with $80\%$ for an observation to be in the training data set, implying that $q=P\{(i,j)\in{\cal S}_{test}\}=0.2$. We used $x_{ij}$ in ${\cal S}_{train}$ to compute $\hat{\bm\theta}_{2,k}$. We then used $\hat{\bm\theta}_{2,k}$ to predict $x_{ij}$ for $(i,j)\in{\cal S}_{test}$. The value of the training-testing cross-validation was $G_{{\cal S}_{test}}^2=\sum_{(i,j)\in {\cal S}_{test}}(x_{ij}-\hat x_{ij})^2$. We called it the mean squares of errors of  prediction (MSEP). For each generated data set, we randomly partitioned ${\cal S}$ into ${\cal S}_{train}$ and ${\cal S}_{test}$ ten times. We used the average of these MSEP values as the estimate of the MSEP value. We derived the estimate of the MSEP value for each candidate $k$. The best $k$ was reported by one with the lowest estimate of the MSEP value. We found that both our proposed and the \textsf{generalizedPCA} methods were able to identify $k$ with almost $100\%$ corrections. The percentages of corrections reported by \textsf{missMDA} were low (Table~\ref{tab:determination of k comparison based on CV}). Based on the estimate of $k$, we compared the root MSEP (RMSEP) values among the three methods (Table~\ref{tab:root prediction mean squares of errors}). Our results showed that the RMSEP values given by our method were always lower than those given by our competitors, indicating that the performance of our method was consistently better than our competitors in both estimation of $k$ and predictions of the missing values.

\begin{table}
\caption{\label{tab:determination of k comparison based on CV}Percentage of correct estimates of $k$ given by training-testing cross-validation in FPCA,  missMDA, and generalizedPCA methods based on simulations with $1000$ replications when data are generated from~\eqref{eq:data generation in simulation}.}
\begin{center}
\begin{tabular}{ccccccccccc}\hline
  & &  \multicolumn{3}{c}{$k$ for FPCA} & \multicolumn{3}{c}{$k$ for missMDA} & \multicolumn{3}{c}{$k$ for generalizedPCA} \\\cline{3-11}
$\tau$ & $n$ & $2$ & $3$ & $4$ & $2$ & $3$ & $4$ & $2$ & $3$ & $4$ \\\hline
$0.1$&$30$&$100.0$&$100.0$&$100.0$&$61.0$&$74.1$&$82.6$&$100.0$&$99.7$&$97.2$\\
&$60$&$100.0$&$100.0$&$100.0$&$79.0$&$85.9$&$91.0$&$100.0$&$100.0$&$100.0$\\
$0.2$&$30$&$99.9$&$99.9$&$100.0$&$62.3$&$76.5$&$87.8$&$99.9$&$98.5$&$92.7$\\
&$60$&$100.0$&$100.0$&$100.0$&$76.9$&$85.4$&$92.1$&$100.0$&$100.0$&$100.0$\\
$0.5$&$30$&$98.0$&$86.7$&$47.1$&$76.7$&$93.4$&$95.3$&$92.8$&$68.6$&$45.3$\\
&$60$&$100.0$&$100.0$&$100.0$&$67.5$&$89.6$&$96.1$&$99.9$&$99.5$&$99.0$\\\hline
\end{tabular}
\end{center}
\end{table}

\begin{table}
\footnotesize
\caption{\label{tab:root prediction mean squares of errors}Estimates of root means squares of errors of prediction (RMSEP) given by $\hat k_{BIC}$ and cross-validation in our FPCA and  cross-validation only in missMDA and generalizedPCA methods from simulations with $1000$ replications when data are generated from~\eqref{eq:data generation in simulation}.}
\begin{center}
\begin{tabular}{cccccccccccccc}\hline
   & & \multicolumn{3}{c}{BIC for FPCA} & \multicolumn{3}{c}{CV for FPCA} &\multicolumn{3}{c}{CV for missMDA} &\multicolumn{3}{c}{CV for generalizedPCA} \\\cline{3-14}
$\tau$ & $n$ & $2$ & $3$ & $4$ & $2$ & $3$ & $4$ & $2$ & $3$ & $4$ & $2$ & $3$ & $4$\\\hline
$0.1$&$30$&$0.273$&$0.262$&$0.250$&$0.273$&$0.262$&$0.250$&$0.291$&$0.277$&$0.264$&$0.607$&$0.730$&$0.864$\\
&$60$&$0.286$&$0.281$&$0.275$&$0.286$&$0.281$&$0.275$&$0.299$&$0.291$&$0.284$&$0.574$&$0.686$&$0.790$\\
$0.1$&$30$&$0.180$&$0.171$&$0.162$&$0.180$&$0.171$&$0.162$&$0.193$&$0.182$&$0.173$&$0.617$&$0.757$&$0.883$\\
&$60$&$0.190$&$0.186$&$0.181$&$0.190$&$0.186$&$0.181$&$0.199$&$0.193$&$0.188$&$0.613$&$0.741$&$0.853$\\
$0.1$&$30$&$0.083$&$0.075$&$0.067$&$0.082$&$0.163$&$0.468$&$0.097$&$0.086$&$0.078$&$0.612$&$0.755$&$0.888$\\
&$60$&$0.092$&$0.088$&$0.084$&$0.092$&$0.088$&$0.084$&$0.099$&$0.094$&$0.090$&$0.667$&$0.795$&$0.898$\\\hline
\end{tabular}
\end{center}
\end{table}

We have developed two approaches to estimate $k$ in our method. To examine which one is better, we compared time taken in the computations. We fount that BIC was more computationally efficient than the training-testing cross-validation. This is because the partition in training-testing cross-validations needs to repeat multiple times to stabilize the results. If the leave-one-out cross-validation were used, then the derivation would be even longer, especially when the size of the data was large.  Therefore, we recommend using $\hat k_{BIC}$ but not the cross-validation to estimate $k$ in practice.

\section{Application}
\label{sec:application}

We apply our method to three real world examples. The first and the second examples are image analysis problems, where we assume that $x_{ij}$ are normally random variables. The third example is count data problem, where we assume that $x_{ij}$ are Poisson random variables. Because all of them belong to exponential family distributions, our method can be applied.

\subsection{Ordinary Image}
\label{subsec:ordinary image}

We applied our method to dimension reduction for ordinary images. Ordinary images are real life images that can be easily obtained by digital cameras or smart phones. Examples include images for street or city views, indoor or outdoor activities, sports, shows, and landscapes. Ordinary images usually contain objects to be recognized. Significance differences can be found by comparing the pixels inside and outside the objects nearby. If traditional PCA is used, then we can only reduce dimensionality of rectangular regions, which cannot avoid the influence of specific objects in the results. This difficulty can be easily addressed by our method.

\begin{figure}
\centering
 \begin{subfigure}[b]{0.35\linewidth}
    \includegraphics[width=\linewidth]{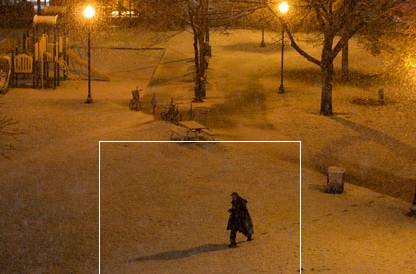}
    \caption{Original Image}
  \end{subfigure}
  \begin{subfigure}[b]{0.35\linewidth}
    \includegraphics[width=\linewidth]{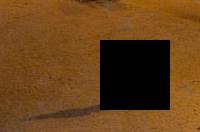}
    \caption{Irregular Shaped Region}
  \end{subfigure}

 \begin{subfigure}[b]{0.35\linewidth}
    \includegraphics[width=\linewidth]{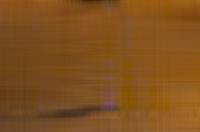}
    \caption{Prediction}
  \end{subfigure}
  \begin{subfigure}[b]{0.35\linewidth}
    \includegraphics[width=\linewidth]{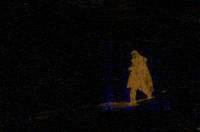}
    \caption{Difference}
  \end{subfigure}
  \caption{Simple FPCA analysis for an irregular shaped portion with the prediction and the absolute difference for the whole rectangular region spanned by the irregular region.}
  \label{fig:an evening image PCA}
\end{figure}

We illustrated our method by its implementation to a street view image after snow at night (Figure~\ref{fig:an evening image PCA}(a)). We used two rectangular windows to scan the image. The larger rectangular window selected pixels to be included. The smaller rectangular window removed pixels selected by the larger window. Therefore, the selected region was irregular. Our method was identical to traditional PCA without the usage of the smaller rectangular window. Traditional PCA could not be applied to the irregular shaped region selected by the two windows. We varied the sizes and centers of the two rectangular windows to select the region. We obtained a number of irregular shaped regions. We applied our method to the regions composed by the pixels inside the larger window but outside the smaller window. We implemented our simple FPCA model developed by~\eqref{eq:modified loglikelihood function of gpca original} to RGB channels individually. We assumed that the distribution given by~\eqref{eq:exponential family distribution} was normal. 

 To implement our method, we first used the penalized likelihood approach to compute $\hat k_{BIC}$ and then use $k=\hat k_{BIC}$ to compute the decomposition. We studied many cases in the choices of the windows. To illustrate our method, it was enough to display our result based on the case displayed by Fgure~\ref{fig:an evening image PCA}(b). In this case, we had $\hat k_{BIC}=4$ for all RGB channels. We used those to derive the final decomposition. The final decomposition was used to recover RGB values for all of the pixels contained by the larger window,  including those contained by the smaller window (Figure~\ref{fig:an evening image PCA}(c)). We looked the absolute difference between the observed and predicted RGB values. It contained the person in smaller window (Figure~\ref{fig:an evening image PCA}(d)). To compare, we also implemented \textsf{missMDA} and \textsf{generalizedPCA} by treating pixels inside the smaller window as missing values. We obtained similar results (not shown). 

\subsection{Medical Image}
\label{subsec:medical image}

\begin{figure}
\centering
 \begin{subfigure}[b]{0.30\linewidth}
    \includegraphics[width=\linewidth]{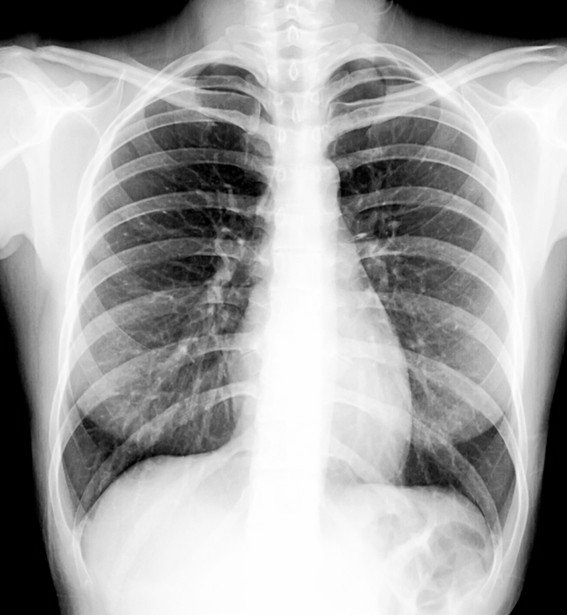}
    \caption{Chest X-Ray}
  \end{subfigure}
  \begin{subfigure}[b]{0.30\linewidth}
    \includegraphics[width=\linewidth]{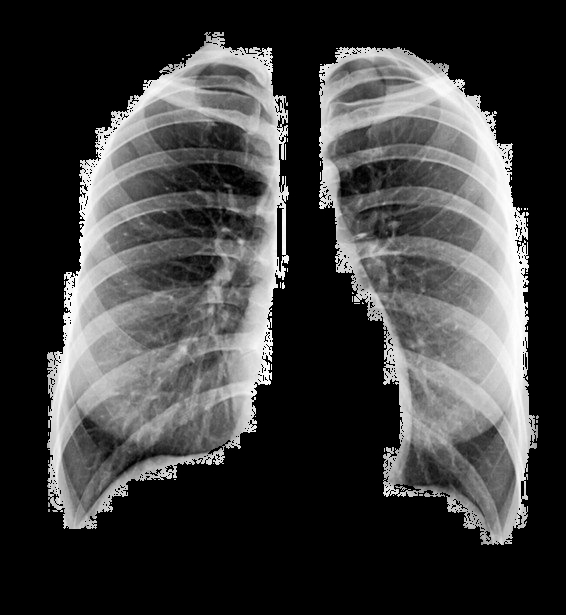}
    \caption{Portion for Lungs}
  \end{subfigure}

 \begin{subfigure}[b]{0.30\linewidth}
    \includegraphics[width=\linewidth]{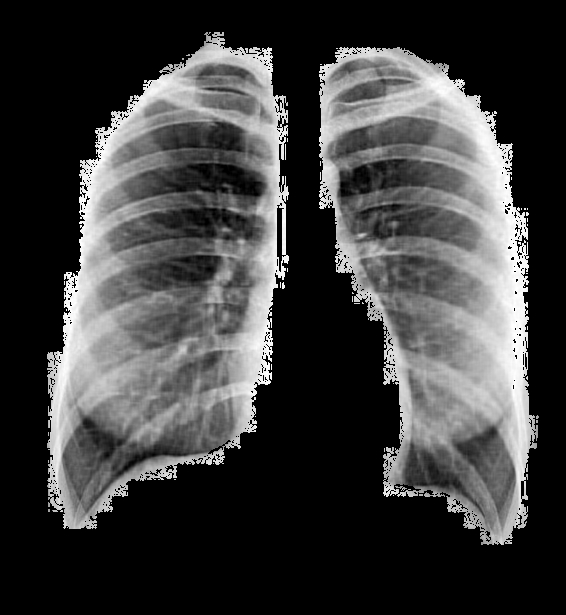}
    \caption{Prediction}
  \end{subfigure}
  \begin{subfigure}[b]{0.30\linewidth}
    \includegraphics[width=\linewidth]{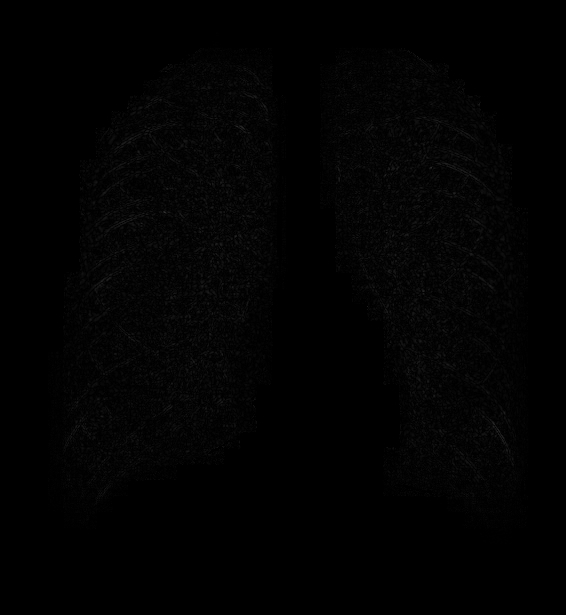}
    \caption{Difference}
  \end{subfigure}
  \caption{Simple FPCA analysis with the prediction and the difference between the observed and predicted values for the lung region.}
  \label{fig:an lung cancer PCA}
\end{figure}

We applied our method to chest X-ray images for lung regions. Chest X-ray images are primarily used by doctors to detect lung diseases, including acute respiratory distress syndrome, pneumonia, tuberculosis, emphysema, and lung cancers. Therefore, the primary interest of chest X-ray images is the lungs. We implemented our method to many chest X-ray images that we have collected. Because their implementations were just replications, it was enough to display the detail of our result based on one of the images. It was the one displayed by Figure~\ref{fig:an lung cancer PCA}, which represented normal lungs.

Our purpose was to reduce the dimensionality of the arbitrary shaped region composed by pixels of lungs. Because the original image did not contain any missing values, it was not a missing data problem. It was inappropriate to include any pixels outside the lungs in dimension reduction because they did not belong to our interest. We removed these pixels from the chest X-ray image and obtained an image for the lungs only (Fgure~\ref{fig:an lung cancer PCA}(b)). Because the RGB values were all identical, it was enough to analyze the information contained by one of the channels. The size of the chest X-ray image was $615\times 556$. The number of the pixels contained by the lungs was $154,\!835$, about $44.5\%$ of the total number of pixels. We assumed that the distribution given by~\eqref{eq:exponential family distribution} was normal.  We implemented the simple FPCA model with the loglikelihood function given by~\eqref{eq:modified loglikelihood function of gpca original} to the lung region. We used BIC to estimate $k$ and obtained $\hat k_{BIC}=38$. We then used $k=38$ to derive the final decomposition. It was used to to recover the image for the lung region only (Figure~\ref{fig:an lung cancer PCA}(c)). We found that it was close to the original image. We confirmed the conclusion by looking at the absolute difference between the recovered image and the original image for the lung region. The result is displayed in Figure~\ref{fig:an lung cancer PCA}(d). 

To compare, we also tried \textsf{missMDA} and \textsf{generalizedPCA}. Their implementations were terminated by error messages without any outputs. The reason probably was that the two methods were designed for missing data, but the study was not a missing data problem. Only our proposed method could be used to this case. It successfully reduced the dimensionality for the irregular shaped region composed by the lungs only. 

\subsection{Dimension Reduction for Count}
\label{secsec:dimension reduction for count}

We applied our method to the country-level daily confirmed COVID-19 data set. The COVID-19 data are reported by the World Health Organized (WHO) from the beginning of the outbreak of COVID-19 (i.e., January 22, 2020) to now (i.e., August 2 2021). The data set contains the numbers of daily new confirmed cases, deaths, and recoveries, given by countries. Because all of the variables are count, a count data set is composed by using the numbers of daily new confirmed cases, deaths or recoveries. Because our interest was the confirmed cases, we excluded deaths and recoveries in the analysis.

Note that FPCA can be implemented to any kinds of exponential family distributions beyond normal. All $x_{ij}$ were count in this implementation and the distribution given by~\eqref{eq:exponential family distribution} represented PMFs for count random variables. We were not able to use traditional PCA or \textsf{missMDA} because they are developed for normally distributed random variables. Therefore, we only need to compare our method with \textsf{generalizedPCA}.

It is well known that the outbreak of COVID-19 has become a worldwide ongoing pandemic since March 2020. The first patient of COVID-19 appeared in Wuhan, China, on December 1 2019. In late December, a cluster of pneumonia cases of unknown cause was reported by local health authorities in Wuhan with clinical presentations greatly resembling viral pneumonia~\cite{chen2020,sun2020}. Deep sequencing analysis from lower respiratory tract samples indicated a novel coronavirus~\cite{feng2020,huang2020}. The virus of COVID-19 primarily spreads between people via respiratory droplets from breathing, coughing, and sneezing, which can cause cluster infections in society. To avoid cluster infections, many countries have imposed travel restrictions in Spring 2020~\cite{zhang2021}. According to the website of WHO, until August 2 2021, the outbreak has affected over 200 countries and territories with more than $200$ million confirmed cases and $4$ million deaths in the entire world. The most serious country is the United States. It has over $36$ million confirmed cases and $632$ thousand deaths. 

\begin{figure}
\centerline{\rotatebox{0}{\psfig{figure=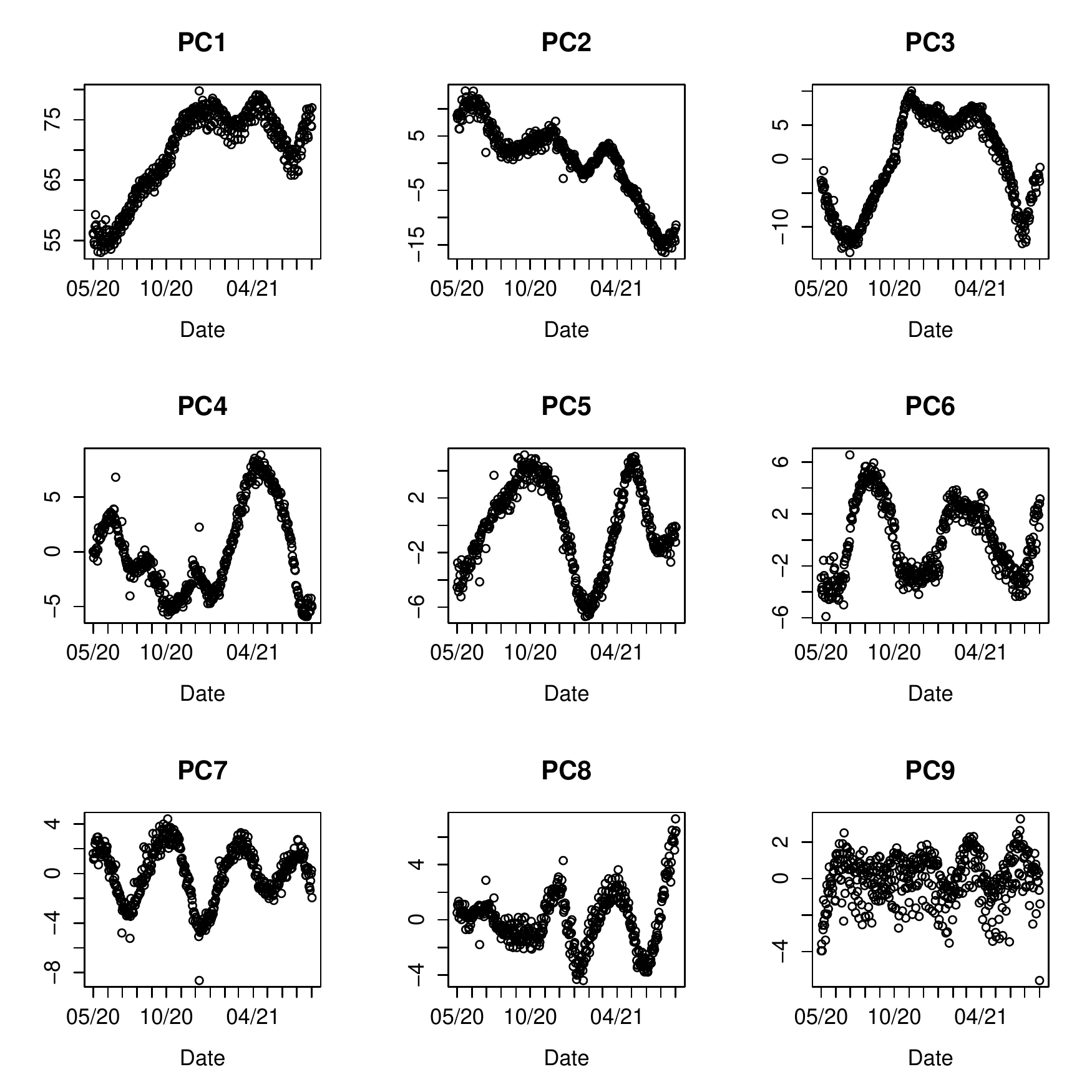,width=3.5in,}}}
\caption{\label{fig:covid-19 PC}PCs in simple FPCA model by assuming that observations follows the quasi-Poisson distribution.}
\end{figure}

We naturally constructed a matrix for count by taking $x_{ij}$ as the number of new confirmed cases with $i$ representing for countries and $j$ for dates. To account for overdispersion, we assumed that $x_{ij}$ followed the quasi-Poisson distribution with dispersion parameter $\phi$. In particular, we assumed that $x_{ij}|\xi_{ij}\sim{\cal P}(\mu_{ij}\xi_{ij})$, where $\xi_{ij}\sim\Gamma(\mu_{ij}/(\phi-1),\mu_{ij}/(\phi-1))$ were independent random effects, such that we had ${\rm E}(x_{ij})=\mu_{ij}$ and ${\rm V}(x_{ij})=\phi\mu_{ij}$. We modeled $\mu_{ij}$ by~\eqref{eq:generalized PCA for exponential family distributions} with $\gamma_{ij}=0$ for all $i$ and $j$,  implying that it was the simple FPCA model. 

\begin{figure}
\centerline{\rotatebox{0}{\psfig{figure=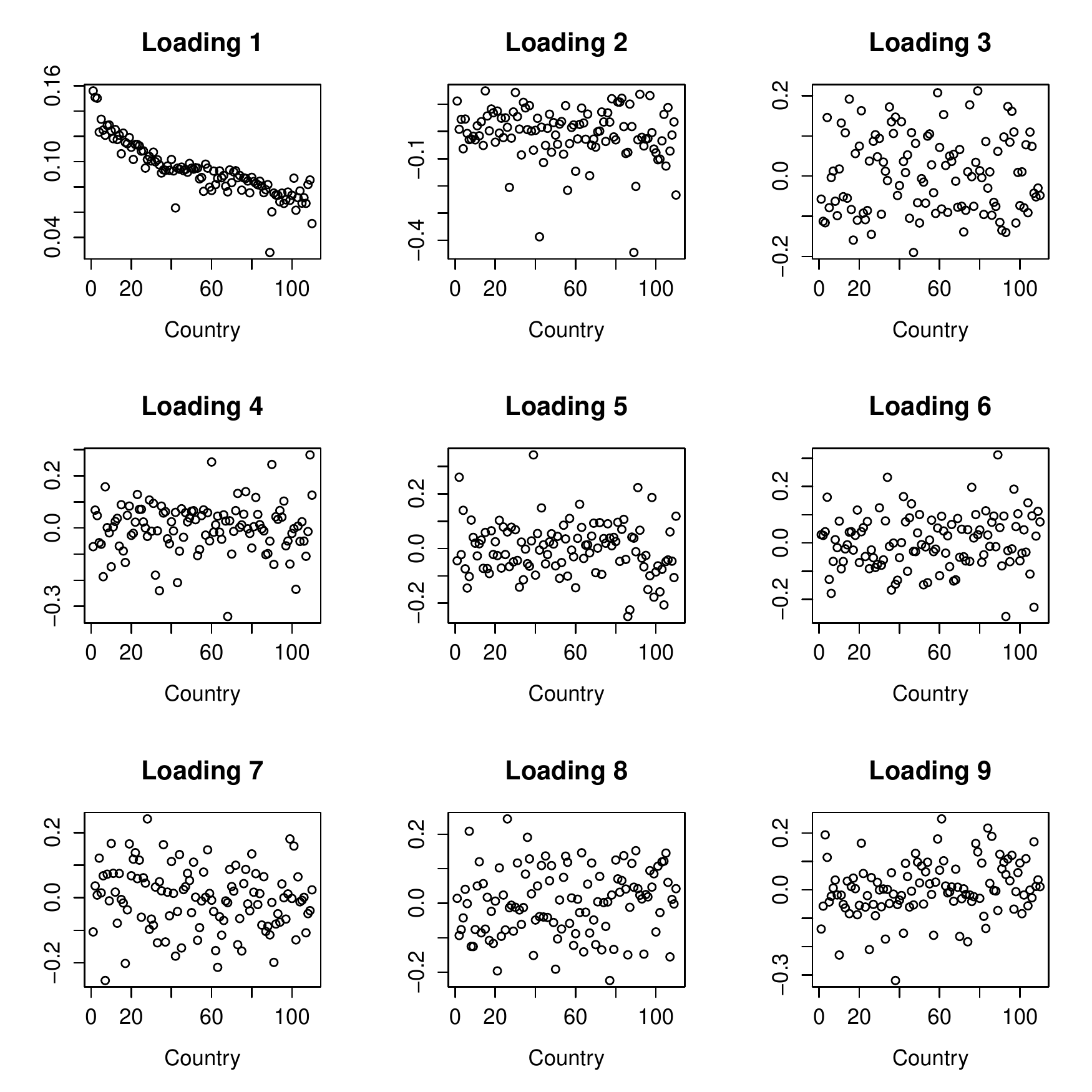,width=3.5in,}}}
\caption{\label{fig:covid-19 loadings}Loadings in simple FPCA model by assuming that observations follows the quasi-Poisson distribution.}
\end{figure}

The data were sparse at the beginning of the outbreak. Many low populated countries reported zeros in many days during the outbreak period. We removed these countries in our analysis. Because China was special, we did not put it in the data. The final data set had $110$ countries and $459$ days. We coded the countries by the decreasing order based on numbers of country-level cumulative confirmed cases until August 2 2021. It covered the period from May 1 2020 to August 2 2021 with many high populated countries, such as the United States (the first country), India (the second country), Brazil (the third country), and etc. The data set contained over $99\%$ of the total number of confirmed cases in the period with more than $90\%$  of the total population in the world (not including China). We estimated $k$ by BIC in the quasi-Poisson model and obtained $\hat k_{BIC}=9$. According to the deviance goodness-of-fit statistic measure ($G^2$), the model accounted for over $98.5\%$ of the total $G^2$ values. We used $k=9$ in~\eqref{eq:modified loglikelihood function of gpca original} and obtained the simple FPCA decomposition. We used it to derive the PCs (Figure~\ref{fig:covid-19 PC}) and the loadings (Figure~\ref{fig:covid-19 loadings}). 

Our result indicated that the first PC accounted for $82.2\%$ total $G^2$ values. It reflected the overall temporal trend of the entire world. It showed that the total number of confirmed cases would increased fast in Fall 2021. Based on the first loading, we claimed that this would occur in the United States, India,  and many European countries. The second PC accounted for $7.85\%$ total $G^2$ values. It contained a decreasing trend in Spring 2020 and 2021 and an increasing trend in Fall 2020. Starting from July 1 2021, the trend became increased with a higher rate than that in Fall 2020, indicating that the outbreak would be more serious in Fall 2021. This trend was also confirmed by the third PCs which account for $3.68\%$ total $G^2$ values. The findings based on the first three PCs were consistent. Based on the trends in Fall 2020, we predicted that the number of daily new confirmed cases would be kept at higher level until the end of 2021. The situation would not change until the beginning of 2022.

We interpreted the first loading by the overall severity of the countries. The pattern was consistent with the over severity of the countries. The vague patterns appeared from the second to the ninth loadings indicated that the infections in different countries were highly correlated with each other. The infections are word-wide. It is impossible to terminate the spread of COVID-19 by countries individually. 

To compare, we also implemented \textsf{generalizedPCA} to the data. Based on the predicted values, we computed its $G^2$ values. Our result showed that \textsf{generalizedPCA} interpreted less $G^2$ values than our method. For instance, if $k=2$, then {generalizedPCA} interpreted about $80\%$ but our simple FPCA interpreted about $90\%$. If $k=9$, then \textsf{generalizedPCA} interpreted about $93.4\%$ but our simple FPCA about $98.8\%$. Therefore, our method better interpreted the overall trends in the numbers of new confirmed cases of COVID-19 than \textsf{generalizedPCA}.

\section{Discussion}
\label{sec:discussion}

In this article, we propose a new dimension reduction method called flexible PCA (FPCA) by extensions of previous PCA models for matrices. An obvious advantage is that our method can be implemented to exponential family distributions in arbitrary shaped regions but traditional PCA can only be implemented to normal distributions in matrices. The research problem investigated by our method is different from that investigated by previous PCA methods for missing values, because we purposely remove observations outside of the region of interest in the analysis even though they are observed. Therefore, missing data mechanism is not an issue in FPCA, but it is an issue in previous PCA methods for missing data. As SVD is a matrix technique, any PCA methods with SVD cannot be implemented to the problem studied by the article. To overcome the difficulty, we propose statistical models with the maximum likelihood approach in the derivation.

 The finding of the article indicates that the usage of statistical models is important in extending matrix-based statistical and machine learning techniques, not limited to traditional PCA for matrices. Matrix data are well-structured and well-organized. Due to rapid development of computer technologies, unstructured or unorganized data are often collected in routine date collection procedures, leading to an important task to extend previous methods, such that they can be used to analyze this kind of data. Our research indicates that statistical models are important in the extensions. Therefore, the impact of the research is not limited to PCA. We believe our idea can also be migrated to other well-known matrix-based methods, such as canonical correlation analysis and factor analysis. This is left to future research. 

\section*{Acknowledgment}
Tonglin Zhang and Baijian Yang's research is supported by NSF-EAGER: SaTC-EDU 10001850; Jing Su's research is partially supported by the Indiana University Precision Health Initiative.

\end{document}